\documentclass{vgtc}                          

\ifpdf
  \pdfoutput=1\relax                   
  \usepackage{graphicx}                
  \DeclareGraphicsExtensions{.pdf,.png,.jpg,.jpeg} 
\else
  \usepackage{graphicx}                
  \DeclareGraphicsExtensions{.eps}     
\fi%

\graphicspath{{figures/}{pictures/}{images/}{./}} 

\usepackage{microtype}                 
\PassOptionsToPackage{warn}{textcomp}  
\usepackage{textcomp}                  
\usepackage{mathptmx}                  
\usepackage{times}                     
\usepackage{cite}                      
\usepackage{tabu}                      
\usepackage{booktabs}                  
\usepackage{subfig}
\usepackage{ccicons} 
\usepackage[dvipsnames]{xcolor}

\onlineid{0}

\vgtccategory{Research}

\vgtcinsertpkg

\title{Talking to Data Visualizations: Opportunities and Challenges}

\author{Gabriela Molina Le\'{o}n\thanks{e-mail: molina@uni-bremen.de}\\ %
        \scriptsize University of Bremen, Germany %
    \and    Petra Isenberg\thanks{e-mail: petra.isenberg@inria.fr}\\ %
     \parbox{1.4in}{\scriptsize \centering Université Paris-Saclay, CNRS, Inria, LISN, France} %
\and Andreas Breiter\thanks{e-mail: abreiter@uni-bremen.de}\\ %
     \scriptsize University of Bremen, Germany}

\teaser{
  \centering
  {\includegraphics[width=\linewidth,keepaspectratio]{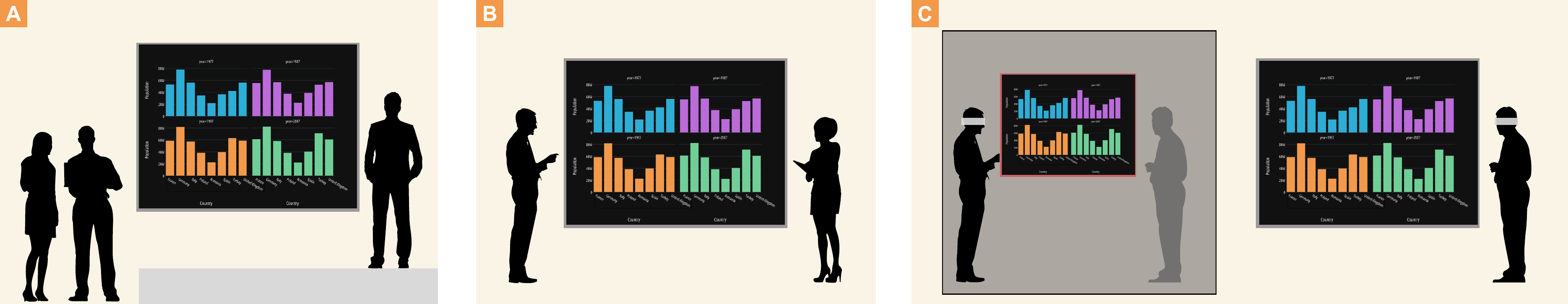}}
  \caption{Examples of collaborative scenarios: (a) A presenter giving a lecture to an audience. (b) Two collaborators interacting with data visualizations. (c) Two persons interact through virtual reality. Silhouettes by 
Gineton Rodrigues~\cite{rodrigues21}, \ccby~CC-BY 4.0.}
  \label{fig:teaser}
}

\abstract{
Speech is one of the interaction modalities that we increasingly come across in natural user interfaces. However, its use in collaborative scenarios has not yet been thoroughly investigated. 
In this reflection statement, we discuss the opportunities and challenges of integrating speech interaction in multimodal solutions for collaborative work with data visualizations. 
We discuss related findings from other research communities and how we could build upon their work to explore and make use of speech interaction for data visualizations in co-located, hybrid, and remote settings. %
} 

\CCScatlist{
  \CCScatTwelve{Human-centered computing}{Interaction design}{}{};
  \CCScatTwelve{Human-centered computing}{Collaborative and social computing systems and tools}{}{};
  \CCScatTwelve{Human-centered computing}{Visu\-al\-iza\-tion}{Visualization design and evaluation methods}{}
}

\begin{document}


\firstsection{Introduction}

\maketitle

Since the launch of ChatGPT in November 2022 \cite{chatgpt}, the interest of the general public in artificial intelligence (AI) has vastly increased \cite{chatgpt23}.
Among the most notorious features of this chatbot is its ability to process questions in more than 50 languages and to generate answers accordingly. As a result, more than 100 million people have interacted with it worldwide so far~\cite{reuters23}.
In the context of such technological advancements, we want to reflect on the opportunities and challenges that leveraging natural language interaction may pose for collaborative data visualization. More specifically, we examine the case where natural language is used via voice input and not only via written text.
Oral communication plays a critical role in collaborative activities, and thus, speech input can leverage people's existing language skills and provide new ways of interacting with data.
In this reflection statement, we highlight the opportunities and challenges that we consider most important for introducing speech interaction into collaborative visual systems.

Most previous work in speech interaction with data visualizations has focused on single-user scenarios~\cite{kim21, srinivasan21}.
We build on this work by discussing how we could transfer findings on speech interaction for single-user activities for designing collaborative systems. Accordingly, we propose a research agenda to understand better how speech interaction could support or influence collaboration. The opportunities and challenges we discuss are by no means exhaustive. Our goal is to start a discussion on the potential role of speech interaction in different collaboration settings.

\section{Speech Interaction}
Speech interaction refers to the use of voice commands to produce a change or response in a computing system. In the field of human-computer interaction, using natural language to interact is recommended due to its expressiveness and the possibility of supporting interaction by broad audiences~\cite{setlur16}. Interacting through written text already works successfully in commercial visualization tools such as Tableau~\cite{tory19}.
Moreover, speech is considered an interaction modality that can help make data visualizations more accessible to, for example, blind and low vision users~\cite{kim23}.
However, speech commands can be hard to discover~\cite{hincapie14}, so using them efficiently requires a learning phase. 
Baughan et al.~\cite{baughan23} analyzed the failures of voice assistants and found that incorrect command execution, missed triggers, and overcapturing affect user trust. 

Prior work on visualization systems for single-user scenarios suggests that the limitations of speech interaction can be addressed by a multimodal interaction design.
Kim et al.~\cite{kim21} designed a mobile application that leverages touch and speech interaction to explore personal health data. In their study, participants found speech fast and flexible to make comparison queries and considered that the combination of touch and speech was helpful to refine previous commands.
Aurisano et al.~\cite{aurisano22} proposed to combine speech and mid-air gestures to create and manipulate views on a wall display and found that their study participants combined both modalities efficiently to create multiple views.
Srinivasan et al.~\cite{srinivasan21} proposed combining touch, pen, and speech interaction to work with unit visualizations. They found that using direct manipulation and speech interaction facilitated a fluid experience and that participants solved speech recognition errors using touch and pen.

For multi-user scenarios, Tse et al.~\cite{tse08} started exploring the multimodal design space for co-located collaboration combining speech and touch in a tabletop application. The authors concluded that if most interaction techniques are associated with speaking, that may influence how often collaborators talk to each other. Additionally, the authors suggested using speech commands for global mode switches and touch gestures for individual mode switches. 
However, speech interaction may have different effects and uses in hybrid and remote settings, where people are not necessarily next to each other, may move around vertical displays, and work asynchronously.

\section{Opportunities}
In the following, we describe the main opportunities that leveraging speech interaction can provide in co-located, hybrid, and remote collaborative work with data visualizations.

\subsection{Interacting from any distance}
Speech input enables distant interaction, which is relevant in collaborative scenarios involving large vertical displays. People tend to navigate physically by walking in front of such screens, and physical navigation may correlate with user performance~\cite{ball07}.
In a study that we conducted recently, participants favored speech over mid-air gestures to interact with data visualizations from afar~\cite{tvcg23}.
While speech can facilitate distant interaction, other modalities can enable interaction up close in a multimodal system. 
For example, Srinivasan et al.~\cite{srinivasan21} proposed to explore network data on large vertical displays combining direct manipulation (touch and pen) with speech interaction.
Such a combination of modalities can be of use not only in co-located scenarios but also in hybrid and remote setups. In those cases, participants may wish to move even while alone in the room (e.g., in a virtual reality scenario as shown in \autoref{fig:teaser}c).

\subsection{Requiring only a microphone}
Regarding hardware, speech interaction only requires a working microphone to listen to the speech commands. Therefore, users are free to use their hands for interacting with the system in other ways. Speech neither requires looking at a screen nor an additional device, so gaze distractions are not a problem.
Regardless of whether the collaboration is happening in a co-located, hybrid, or remote setting, the physical characteristics of the interaction do not change. 
Additionally, using a microphone is a standard in collaborative commercial platforms, especially when the collaborators need to communicate synchronously. Thus, no extra equipment is necessary. Giving each person access to a microphone in a co-located scenario, however, is recommended to ensure accuracy and to identify each speaker.
From a software perspective, speech recognition only demands using dedicated libraries or tools, such as web APIs (e.g., Mozilla's API~\cite{speechapi}) or AI-based tools to support a custom vocabulary (e.g.,~Picovoice \cite{picovoice}). Nowadays, being an AI expert is not a requirement to leverage AI-based approaches. While free and open-source toolkits are still rare, there are already a few options, such as Vosk \cite{vosk}.

\subsection{Adapting to speech-heavy activities}
Collaborative scenarios can involve not only sensemaking but also other activities, such as presenting and teaching about visualization techniques and tools (e.g., \autoref{fig:teaser}a), where one person is responsible for most of the oral communication.
In such cases, we could incorporate speech commands directly into the presentation or teaching content. The presenter or the audience can then use other modalities to interact while the person is talking. Interactive speeches could work in co-located, hybrid, or remote settings, as long as the setup includes a microphone.
Although voice assistants are usually activated with a wake word (e.g., ``Alexa'', ``Hey Siri'', etc.), using such a phrase is not required. Instead, the system could listen to the conversation, waiting for an opportunity to participate~\cite{zargham22}, for example, to provide a data fact relevant to the discussion. Studying storytelling techniques may help to propose ways of incorporating speech commands into a presentation or lecture. For example, Shin et al.~\cite{shin23} proposed a related solution by generating natural language narratives to present a sequence of data changes. Their system created textual narratives and animations to highlight temporal changes in a scatterplot. Thus, the presenter could trigger the animations with commands that form parts of their speech and use other interaction modalities for further effects. 

\subsection{Supporting multilingual interaction}
Nowadays, speech interaction tools support the recognition of dozens of languages, which could support collaborative work between people who speak different languages.
Previous work shows that people appreciate speech interaction because they can refer orally to concepts that are tedious to specify as data queries through direct manipulation \cite{setlur16}. The option to interact orally in their native language may improve the user experience of the collaborators.
Moreover, a system could provide similar flexibility for using expert and non-expert terms to interact orally. For example, a person could say ``draw a line that shows if my points are increasing or decreasing'' while someone else could say ``add a linear trend line''.

\section{Challenges}
We now present the main challenges that researchers and practitioners may face when leveraging speech interaction in the design of collaborative systems.

\subsection{Recognition errors}
In the context of interactive data visualization, Srinivasan and Stasko \cite{srinivasan18} reported that study participants sometimes became frustrated while interacting due to speech detection errors. When voice assistants fail to understand speech commands time after time, people tend to trust them less \cite{baughan23}. 
Providing multimodal interaction may be a solution to this challenge. For example, in the study of Saktheeswaran et al.~\cite{saktheeswaran20}, participants appreciated being able to correct speech recognition and ambiguity errors via touch.
Moreover, we are currently witnessing significant improvements in voice recognition systems. The recent launch of deep-learning-based tools such as Whisper \cite{radford23}, Vosk \cite{vosk}, and Picovoice \cite{picovoice} suggests promising advances in speech interaction accuracy.

\subsection{Collaboration conflicts}
Given that dialogue is a fundamental aspect of collaboration, using speech to interact with the system, and not only with other humans, may disturb the conversation flow. Collaboration partners may hesitate to use speech commands to avoid interrupting each other.  
However, in our recent elicitation study~\cite{tvcg23}, participants sometimes wished for the system to listen to them and directly map part of their conversation to interactions with the data visualizations.
Furthermore, problems may arise during turn-taking, as overlapping speech commands may make speech recognition difficult. As the number of collaborators increases, avoiding overlapping voices becomes harder. In this situation, it is also crucial that the system gives appropriate feedback to help the collaborators understand whether the speech commands are understood and processed.

\subsection{Privacy concerns}
People may not necessarily be comfortable with a device constantly listening to them, waiting for voice input. Although the wish for privacy-friendly designs can depend on the country and social norms~\cite{seiderer20}, privacy is a primary factor for user acceptance. Moreover, people are protective of saving and making voice recordings of themselves and others available online with cloud services they do not trust. Thus, it is necessary to find alternatives to constant listening and storing voice recordings~\cite{malkin19}. 
Additionally, speech interaction requires talking out loud, which does not allow users to make their interactions private. However, that may not be a problem in asynchronous collaborations. Also, multimodal systems could support private interactions through other modalities to complement speech-based public interactions~\cite{tse08}. For limiting data sharing, there are speech tools that do not require processing voice input online (e.g., Picovoice~\cite{picovoice}). They process it on the local machine, which may be a suitable alternative.

\subsection{Support for language learners}
Although chatbots and voice assistants support multiple languages, we must consider that many people interact with computing systems in English or another language that may not be their first language. Having an accent or not being familiar with the pronunciation of specific words can be an obstacle to speech recognition~\cite{srinivasan18, molina22}. Similarly, there are thousands of spoken languages worldwide, and thus, speech interaction is limited by the languages supported by the corresponding software. Machine translation advances may provide more support in the future, but it has limits as well.

\section{Research Agenda}
Based on the opportunities and challenges mentioned above, we list the research directions we consider most critical and promising for developing scientific knowledge on the role of speech interaction in multimodal and collaborative scenarios.

\subsection{Understanding speech in collaboration}
As research on conversational user interfaces and multimodal interaction has focused on single-user scenarios, we need to conduct more studies to investigate different collaborative activities to understand how speech interaction can support and affect collaborative work with data visualizations.
Similar to how Kim et al.~\cite{kim23} created and published a corpus of questions that participants posed in their study for future speech-based design, collaboration studies could help generate such corpora of questions and dialogues that may be relevant for designing collaborative systems. 
Collecting these data can help us understand better what tasks are most suitable to perform via speech during collaborative work and what commands are relevant for different collaboration settings and styles.

\subsection{Leveraging more design methods}
Although current commercial products supporting natural language have known limitations in speech recognition, that should not be an impediment to investigating the potential of speech interaction.
Given the latest advancements in AI, natural language technologies may evolve rapidly.
We should think beyond the current technological capabilities, as diverse research communities have successfully done. Voice assistant researchers often investigate and propose conversational scenarios through storyboards and Wizard of Oz studies (e.g.,~\cite{zargham22, kernan22}) to investigate the relevant factors in conversational interface design. We should leverage such methods to inform the design of future visual systems by exploring what could help collaborators most and how.

\subsection{Exploring collaborative and multimodal solutions}
As mentioned above through different examples, many limitations of speech interaction can be addressed by combining speech with other interaction modalities. Prior work suggests that combining natural language with direct manipulation is a promising solution~\cite{srinivasan21}, as well as combining speech commands with mid-air gestures~\cite{aurisano22}. We should examine how collaboration partners use multiple interaction modalities and how those may influence their strategies and interaction choices. Combining modalities that support up close and distant interaction could help support collaboration at different distances from the visualizations~\cite{tvcg23}.

\subsection{Providing voice-based feedback}
Natural language interfaces can include not only voice input but also voice output. Given the advancements in voice assistants, we should explore the scenarios in which a system can generate prompts in natural language and be part of the conversation.
Zargham et al.~\cite{zargham22} already started exploring these scenarios by investigating in what situations a proactive voice assistant could intervene in a decision-making conversation or debate. In the context of presentations and teaching activities, we should consider how a voice assistant could support explanatory visualizations or how it could explain how to read a visualization.
Hence, it is worth investigating how the human-assistant interaction could work and how the participation of voice assistants may influence the collaborators. This would be an interdisciplinary endeavour to better understand flaws in communication. 

\section{Conclusion}
This reflection statement discussed the use of speech interaction in collaborative scenarios.
We reflected on the opportunities and challenges of incorporating this interaction modality into multi-user interactive systems. 
Accordingly, we discussed how to leverage speech in diverse collaborative activities and setups involving data visualizations.
Given the recent advancements in AI-based approaches for natural language interaction, we should investigate how natural language could facilitate working with data visualizations from any distance, in a group, and even in multiple languages.
We should explore this topic from the different perspectives of the related research communities, such as those working on human-computer interaction, natural language processing, and computer-supported cooperative work.
Such interdisciplinary efforts will help us better understand whether and how speech could become a valuable interaction modality in collaborative visual systems.

\acknowledgments{
This work was partly funded by the Deutsche Forschungsgemeinschaft (DFG, German Research Foundation) -- project number 374666841 -- SFB 1342.}

\bibliographystyle{abbrv-doi-hyperref-narrow}

\bibliography{template}
\end{document}